\documentclass[aps,prl,twocolumn,superscriptaddress,showpacs]{revtex4}
\usepackage[utf8]{inputenc}
\usepackage{indentfirst}
\usepackage{amsmath}
\usepackage{amssymb}
\usepackage{amsthm}
\usepackage{mathtools}
\usepackage{dpackage,hyperref}
\usepackage{xcolor,ulem}

\hypersetup{
  colorlinks   = true, 
  urlcolor     = red, 
  linkcolor    = blue, 
  citecolor   = violet 
} 

\newcommand{\rmi}{\mathrm{i}}
\newcommand{\rme}{\mathrm{e}}

\begin{document}
\title{Distributed Quantum Metrology with a Single Squeezed-Vacuum Source}

\author{Dario Gatto\footnote{dario.gatto@port.ac.uk}}
\affiliation{School of Mathematics and Physics, University of Portsmouth, Portsmouth PO1 3QL, United Kingdom}

\author{Paolo Facchi}
\affiliation{Dipartimento di Fisica and MECENAS, Universit\`a di Bari, I-70126  Bari, Italy}
\affiliation{INFN, Sezione di Bari, I-70126 Bari, Italy}

\author{Frank Narducci}
\affiliation{Department of Physics, Naval Postgraduate School, Monterey, California 93943}

\author{Vincenzo Tamma\footnote{vincenzo.tamma@port.ac.uk}}
\affiliation{School of Mathematics and Physics, University of Portsmouth, Portsmouth PO1 3QL, United Kingdom}
\affiliation{Institute of Cosmology and Gravitation, University of Portsmouth, Portsmouth PO1 3FX, United Kingdom}

\date{\today}
\begin{abstract}
We propose an interferometric scheme for the estimation of a linear combination with non-negative weights of an arbitrary number $M>1$ of unknown phase delays, distributed across an $M$-channel linear optical network, with Heisenberg-limited sensitivity. This is achieved without the need of any sources of photon-number or entangled states, photon-number resolving detectors or auxiliary interferometric channels. Indeed, the proposed protocol remarkably relys upon a single squeezed-state source, an antisqueezing operation at the interferometer output, and on-off photodetectors.
\end{abstract}

\pacs{%
03.65.Wj, 
42.50.St, 
06.20.-f, 
03.67.-a
}

\maketitle

\paragraph{Introduction and Motivations.}
Quantum metrology aims at harnessing inherently quantum features such as entanglement, multi-photon interference and squeezing, to develop novel quantum enhanced technologies for sensing and imaging beyond any classical capabilities~\cite{caves81,yurke86,holland93,giovannetti04,giovannetti06,giovannetti11,pezze14,demkowicz15}. 
More recently a great deal of attention has been devoted to distributed quantum metrology, particularly on the problem of measuring a linear combination of several unknown phase shifts distributed over a linear optical network~\cite{eldredge18,proctor17}. More explicitly, we will be interested in measuring a linear combination of $M>1$ unknown distributed phases. This problem is of interest in a variety of settings: from the mapping of inhomogenous magnetic fields~\cite{steinert10,hall12,pham11,seo07,baumgratz16}, phase imaging~\cite{humphreys13,liu16,yue14,knot16,gagatsos16,ciampini15} and quantum-enhanced nanoscale nuclear magnetic resonance imaging~\cite{eldredge18,arai15,lazariev15}, to applications in precision clocks~\cite{komar14}, geodesy, and geophysics~\cite{nabighian05,wright04,stacey64,hamalainen93}.

A novel scheme was recently proposed to tackle distributed quantum metrology with Heisenberg limited sensitivity~\cite{ge17}. However, its main limitation is the fact that it relies on two Fock states with a large number of photons  as probes in order to achieve Heisenberg-limited sensitivity. Schemes to make high photon number Fock states do not currently exist. Furthermore, it requires a number of auxiliary interferometric channels up to the number $M$ of unknown distributed phases, and photon-number-resolving detectors. Therefore, devising measurement schemes which can exhibit supersensitivity while making use of probe states which are simple to produce in the laboratory with current technology is a matter of great interest.

\begin{figure}
\includegraphics[width=.49\textwidth]{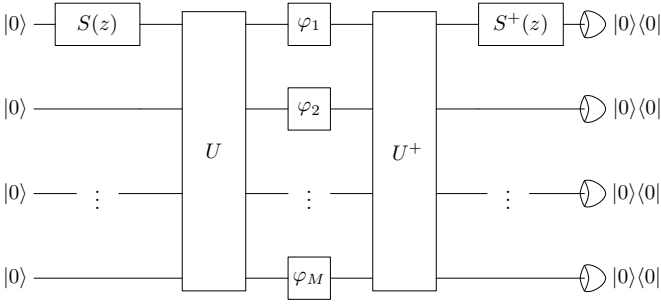}
\caption{Interferometric setup with only a single squeezed-vacuum source for Heisenberg limited estimation of the linear combination of unknown phases $\varphi_i$, with $i= 1,...,M$, as in Eq.~\eqref{eq:qdef}. The linear optical networks represented by $\hat{U},\hat{U}^\dag$ are set up in such a way to satisfy Eq.~\eqref{eq:encoding}. The operators $\hat{S}(z),\hat{S}^\dag(z)$ represent squeezing and anti-squeezing operation, respectively.}
\label{fig:circuit} 
\end{figure}

We overcome these limitations introducing an interferometric scheme (Fig.~\ref{fig:circuit}) which employs only a single squeezed source and on-off photodetectors. Indeed, squeezed states of light are a natural candidate for Heisenberg-limited probing~\cite{manceau17,maccone19}, on account of their experimental availability with high mean photon number and their non-classical character. While such states have been largely used to yield supersensitivity in the estimation of a single unknown parameter, their quantum metrological advantage in the case of multiple distributed parameters has not yet been fully explored~\cite{anisimov10}. Here, we demonstrate how a simple $M$-channel linear optical interferometer with only a single squeezed-vacuum source and on-off photodetectors can achieve Heisenberg-limited sensitivity in distributed quantum metrology with $M$ unknown phase delays. Remarkably, such a scheme can be implemented experimentally with present quantum optical technologies.

\paragraph{The Optical Interferometer.}
We describe here in details an interferometric setup (Fig.~\ref{fig:circuit}) able to estimate the combination
\begin{equation}
\overline{\varphi} = \sum_{j=1}^Mw_j\varphi_j,
\label{eq:qdef}
\end{equation}
of $M$ unknown phases $\varphi_j$ ($j=1,...,M$) for any given set of non-negative weights $\{w_j\}_{j=1}^M$~\cite{nota1}. 
Without loss of generality we will assume in the following the normalization 
$\sum_j w_j =1$, so that the $w_j$ are probability weights. The general situation will differ just by an immaterial  factor.

The probe light at the input of our interferometer is prepared in the squeezed-vacuum state
\begin{equation}
\ket{\Psi_{\rm in}} 
= \hat{S}_1(z) \ket{\Omega} ,
\end{equation}
where $\ket{\Omega}=\ket{0}_1\cdots \ket{0}_M$ is the vacuum state,
$\hat{S}_1(z)=\rme^{{1\over2}(z^*\hat{a}_1^2-z \hat{a}_1^{\dag2})}$ is the squeezing operator, $\hat{a}_1$ is the photonic annihilation operator of the first mode, and $z$ is the squeezing parameter. 
The squeezing parameter $z$ fixes the mean number  of input photons $\langle \hat{N}\rangle = \bra{\Psi_{\rm in}} \hat{N} \ket{\Psi_{\rm in}}$, with $\hat{N}= \sum_j \hat{a}_j^\dagger\hat{a}_j$, by the relation $\langle \hat{N}\rangle=\bar{N}$, where 
\begin{equation}
\qquad  \bar{N}= \sinh^2(|z|).
\label{eq:meanN}
\end{equation}

The probe travels through the first linear optical transformation, described by the unitary operator $\hat{U}$ through the equation
\begin{equation}
\hat{U}^\dag \hat{a}_i\hat{U} = \sum_{j=1}^M{\cal U}_{ij}\hat{a}_j,
\label{eq:passive_linear_optical}
\end{equation}
where ${\cal U}$ is an $M\times M$ unitary matrix associated with the transition amplitudes from the channel $j$ to the channel $i$, with $i,j=1,\dots,M$. We encode the weights of Eq.~\eqref{eq:qdef} in the matrix elements of $\cal U$ via
\begin{equation}
\mathcal{U}_{j 1} = \sqrt{w_j},
\label{eq:encoding}
\end{equation}
with $j=1,\dots,M$.
This can always be achieved with an appropriate combination of beam splitters~\cite{reck94}, and more importantly, creates entanglement~\cite{ge17} in the squeezed state, which is distributed across all channels containing the phase delays $\varphi_1,\dots,\varphi_M$. 

After the linear optical transformation $\hat{U}$ the probe undergoes phase shifts $\varphi_1,...,\varphi_M$ through the respective channels, and finally evolves through the inverse linear optical transformation $\hat{U}^\dag$. Reversing the linear optical transformation will allow us to effectively project the output state onto the input state (see the next paragraph).
Thus, given the generator of the phase shifts,
\begin{equation}
\hat{G} = \sum_{j=1}^M\varphi_j \hat{a}^\dag_j \hat{a}_j,
\label{eq:Gdef}
\end{equation}
the state at the output of our interferometer is
\begin{equation}
\ket{\Psi_{\rm out}} = \hat{U}^\dagger \rme^{-\rmi \hat{G}}\hat{U}\ket{\Psi_{\rm in}}.
\end{equation}

\paragraph{Heisenberg-Limited Estimation.} We now demonstrate Heisenberg-limited sensitivity (in Eq.~\eqref{eq:sensitivity}) by means of the observable 
\begin{equation}
\hat{{\cal O}}=\ket{\Psi_{\rm in}}\bra{\Psi_{\rm in}}, 
\end{equation}
associated to the projection of the output state over the input state, i.e. to the probability that the probe leaves the interferometer with its state unaltered. Since the expectation value of $\hat{\cal O}$ is
\begin{align}
\langle \hat{{\cal O}}\rangle_{\mathrm{out}}  &=  \langle{\Psi_{\rm out}}|\hat{{\cal O}}|{\Psi_{\rm out}}\rangle = |\bra{\Psi_{\rm in}}\hat{U}^\dagger \rme^{-\rmi \hat{G}} \hat{U}\ket{\Psi_{\rm in}}|^2 
 \notag\\
 &=|\bra{\Omega}\hat{S}_1^\dag(z)\ket{\Psi_{\rm out}}|^2,
  \label{eq:expvaldef}
\end{align}
the measurement of $\hat{{\cal O}}$ is equivalent to projecting onto the vacuum $\ket{\Omega}$ after the action of an anti-squeezing operation on the first channel, described by $\hat{S}_1^\dag(z)$. This can be experimentally achieved, for instance, by retro-reflecting the down-converted photons onto the crystal generating the original squeezed light~\cite{resch01,resch02,resch02-2,lundeen09,demkowicz12}, and then using on-off photodetectors.

Since $\langle \hat{\mathcal{O}} \rangle_{\mathrm{out}} = |\langle{\Psi_{\rm in}}|{\Psi_{\rm out}}\rangle|^2$ is the probability of the output state to coincide with the input state, if the phases are small, the total interferometric operator should be close to the identity, and therefore $\langle \hat{\mathcal{O}} \rangle_{\mathrm{out}}$ should be close to one. More precisely, since $-|\varphi|_{\mathrm{max}} \hat{N}\leq \hat{G} \leq |\varphi|_{\mathrm{max}} \hat{N}$, 
with $|\varphi|_{\mathrm{max}} = \max_i |\varphi_i|$, 
and the interferometer preserves the total number of photons, if 
\begin{equation}
|\varphi|_{\mathrm{max}}  \langle \hat{N}\rangle  \ll 1,
\label{eq:phases_expansion_condition}
\end{equation}
we can perform an expansion of $\langle \hat{\mathcal{O}} \rangle_{\mathrm{out}}$ in powers of $\hat{G}$.

By using the notation $\langle \hat{G}^m\rangle_U$ for the expectation value of the operator $\hat{G}^m$ with $m=1,2$ taken at the state $\ket{\Psi_U} = \hat{U}\ket{\Psi_{\rm in}}$, and $\Delta G_U^2 = \langle \hat{G}^2 \rangle_U - \langle \hat{G} \rangle_U^2$, for the variance of $\hat{G}$, we obtain
\begin{eqnarray}
\langle\hat{\mathcal{O}} \rangle_{\mathrm{out}} &=& \bigl|\bigl\langle \rme^{-\rmi \hat{G}} \bigr\rangle_U \bigr|^2 
\simeq \Bigl|\Bigl\langle 1 - \rmi \hat{G} - \frac{1}{2} \hat{G}^2 \Bigr\rangle_U \Bigr|^2
\nonumber\\
&\simeq& 1 -  \Delta G_U^2,
\label{eq:expansion}
\end{eqnarray}
up to fourth-order terms~\cite{nota2}.

By using Eq.~\eqref{eq:Gdef} and the canonical commutation relations~\cite{nota3},we obtain that the exact expression for the variance of $\hat{G}$ depends on $\overline{\varphi}$ in Eq.~\eqref{eq:qdef}, and on $\overline{\varphi^2} = \sum_j w_j\varphi_j^2$ as
\begin{equation}
\Delta G^2_U = \overline{\varphi}^2 \, \bigl(\langle \hat{N}^2 \rangle -  \langle \hat{N} \rangle^2 \bigr) + \bigl(\overline{\varphi^2} - \overline{\varphi}^2\bigr)\, \langle \hat{N} \rangle ,
\end{equation}
where $\langle \hat{N}^2 \rangle = \bra{\Psi_{\mathrm{in}}} \hat{N}^2 \ket{\Psi_{\mathrm{in}}}$. The variance of $\hat{G}$
is made of a contribution from  number fluctuations and a contribution from the fluctuations of the phases $\varphi_j$ with respect to the weights $w_j$.

This result is valid for any (not necessarily Gaussian) $M$-boson state $\ket{\Psi_{\mathrm{in}}}$ with all modes but the first in the vacuum.
In our case, since the first mode is in a squeezed vacuum state, its photon-number statistics is super-Poissonian~\cite{teich89}, with mean photon number $\langle \hat{N} \rangle  =\bar{N}$ given by~\eqref{eq:meanN}
and a variance
\begin{equation}
\langle \hat{N}^2 \rangle -  \langle \hat{N} \rangle^2 = 2 \bar{N} (\bar{N} +1),
\end{equation} 
which scales as $\bar{N}^2$. This scaling is unlike a coherent state which has a Poissonian photon-number statistics with variance equal to the mean $\langle \hat{N}\rangle$. As we will see, this is an essential ingredient for obtaining a Heisenberg-limited sensitivity.

For large $\bar{N}$ 
one gets $\Delta G^2_U \simeq 2\bar{N}^2 \overline{\varphi}^2$,
whence
the expectation value of our observable~\eqref{eq:expansion} reads
\begin{equation}
\langle \hat{\mathcal{O}}\rangle_{\mathrm{out}} \simeq 1 - {2\bar{N}^2 \overline{\varphi}^2},
\label{eq:expval}
\end{equation}
and differs from 1 by a small quantity, as expected. Indeed, we are in the regime of large $\bar{N}$ and small 
$\overline{\varphi}$, such that  $| \overline{\varphi} | \bar{N} \leq |\varphi|_{\mathrm{max}}  \bar{N}$
is small, in accordance with Eq.~\eqref{eq:phases_expansion_condition}. 

The sensitivity in the estimation of $\overline{\varphi}$ is obtained by the error propagation formula~\cite{pezze14}
\begin{equation}
\delta \varphi^2 = \frac{\langle \hat{\mathcal{O}}^2 \rangle_{\mathrm{out}} -\langle \hat{\mathcal{O}} \rangle_{\mathrm{out}}^2}
{\bigl({\diff\over\diff q}\langle \hat{\mathcal{O}}\rangle_{\mathrm{out}} \bigr)^2},
\end{equation}
By using the fact that $\hat{{\cal O}}=\hat{{\cal O}}^2$ is a projection, and by virtue of Eq.~\eqref{eq:expval},
we easily get
\begin{align}
\delta \varphi^2 &\simeq {1\over8 \bar{N}^2},
\label{eq:sensitivity}
\end{align}
i.e. the sensitivity scales at the Heisenberg limit. 

\begin{figure}
\includegraphics[width=.49\textwidth]{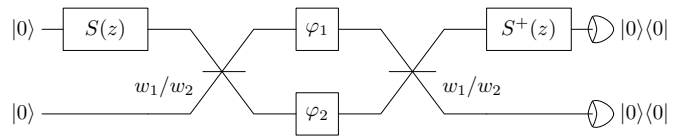}
 \caption{Estimation of the linear combination two unknown phases $\varphi_1,\varphi_2$ with weights $w_1,w_2$. The interferometric scheme in Fig.~\ref{fig:circuit} reduces to a Mach-Zehnder interferometer, with unbalanced beam splitters with $R/T = w_1/w_2$, where $R$ is the reflectivity and $T$ is the trasmittivity.}
 \label{fig:experimental_schematic}
\end{figure}

\paragraph{Example.} Let us consider the $M=2$ case, i.e. we wish to estimate
\begin{equation}
 \overline{\varphi} = w_1\varphi_1+w_2\varphi_2,
\end{equation}
in a 2-mode interferometer, for assigned  weights $w_1,w_2\ge0$, $w_1+w_2=1$. One possible choice for $\cal U$ which satisfies Eq.~\eqref{eq:encoding} is
\begin{equation}
{\cal U} = 
\begin{pmatrix}
                             \sqrt{w_1} & \sqrt{w_2} \\
                             \sqrt{w_2} & -\sqrt{w_1}
                            \end{pmatrix}.
\end{equation}
This is just the matrix describing a beam splitter of reflectivity 
$R=w_1$ and transmittivity $T=w_2$, 
therefore the interferometric setup is simply that of a Mach-Zehnder interferometer, see Fig.~\ref{fig:experimental_schematic}. Remarkably, here both phases $\varphi_1$ and $\varphi_2$ are unknown, differently from previous proposals where only one parameter is unknown~\cite{caves81,takeoka17}. 

Even more interestingly, the scheme in Fig.~\ref{fig:experimental_schematic} is sensitive to the \textit{sum}, rather than the difference, of the phases $\varphi_1$ and $\varphi_2$ with positive weights $w_1$ and $w_2$, respectively. To see how this is possible, let us set $w_1=w_2=1/2$ for simplicity, and let us consider the optical unitary transformation describing the balanced Mach-Zehnder,
\begin{equation}
 \hat{U}_{MZ} = \rme^{\frac{\rmi}{2}(\varphi_1-\varphi_2)\hat{J}_y}\rme^{-\frac{\rmi}{2}(\varphi_1+\varphi_2)\hat{N}},
 \label{eq:MZ_unitary}
\end{equation}
where $\hat{J}_y=-\frac{i}{2}(\hat{a}^\dag_1\hat{a}_2-\hat{a}_1\hat{a}^\dag_2)$~\cite{yurke86}. As we can see the output state $\ket{\Psi_{\rm out}}=\hat{U}_{MZ}\ket{\Psi_{\rm in}}$ does depend, in general, on both $\varphi_1$ and $\varphi_2$, however the information on the sum of the phases can be ``washed out'' by the choice of the measurement protocol. Indeed, if $\ket{\Psi_{\rm in}}$ is an eigenstate of the number operator $\hat{N}$, $\ket{\Psi_{\rm out}}$ depends \textit{only} on the relative phase $\varphi_1-\varphi_2$, because the second exponential in Eq.~\eqref{eq:MZ_unitary} gives rise to a global complex phase, and the information on $\overline{\varphi}=(\varphi_1+\varphi_2)/2$ is completely lost. Furthermore, if one measures an observable $\hat{{\cal O}}$ which commutes with $\hat{N}$, then $\langle\hat{{\cal O}}\rangle_{\rm out}$, as well as all the higher moments $\langle\hat{{\cal O}}^k\rangle_{\rm out}$~\cite{yurke86}, will again depend on $\varphi_1-\varphi_2$ only, since $\hat{U}^\dag_{MZ}\,\hat{{\cal O}}\,\hat{U}_{MZ} = \rme^{-\frac{\rmi}{2}(\varphi_1-\varphi_2)\hat{J}_y}\,\hat{\mathcal{O}}\,\rme^{\frac{\rmi}{2}(\varphi_1-\varphi_2)\hat{J}_y}$.

\paragraph{Discussion.}
We have shown how squeezed light can be used to estimate an arbitrary superposition of phases with non-negative weights. Our protocol can overcome the limitations of~\cite{ge17}, most notably we can achieve the Heisenberg limit with a single squeezed state rather than two Fock states. Futhermore, our protocol does not necessitate the use of auxiliary channels nor photon-number-resolving detectors. The interferometric setup is easily realizable for any set of weights by using only beam splitters and phase shifters~\cite{reck94}. The initially separable input state acquires entanglement across the various channels where the phase shifts are distributed thanks to the linear optical network. The squeezed source can be produced in a number of ways, including spontaneous parametric down-conversion and four-wave mixing. Anti-squeezing has already been achieved with high efficiency~\cite{resch01,resch02,resch02-2,lundeen09,demkowicz12} by retro-reflecting the down-converted photons and the pump back onto the crystal. We would like to mention that it is also possible to have, instead of the antisqueezer $\hat{S}^\dag(z)=\hat{S}(-z)$, an output squeezer $\hat{S}(z')$ where $|z'|\neq|z|$, but $z$ and $z'$ have opposite complex phases. Analysis of this sort of interferometer protocol is beyond the scope of this letter. Remarkably, given the parameter of the first squeezer, increasing the parameter of the second one can compensate for detection losses~\cite{manceau17}. In conclusion, our protocol can achieve Heisenberg-limited sensitivity for distributed quantum metrology while being well within the realm of current quantum optical technologies.

\paragraph{Acknowledgements.}  This project was partially supported by the Office of Naval Research Global (N62909-18-1-2153). DG is supported by the University of Portsmouth. PF is partially supported by Istituto Nazionale di Fisica Nucleare (INFN) through the project ``QUANTUM", and by the Italian National Group of Mathematical Physics (GNFM-INdAM).

\begin{widetext}
\section{Supplemental Material}
\subsection{Expansion of the Observable Expectation Value}
\label{sec:series_expansion}
Let $\hat{G}$ be the generator of the phase shifts, i.e.
\begin{equation}
\hat{G} = \sum_{j=1}^M\varphi_j \hat{a}^\dag_j \hat{a}_j.
\end{equation}
If $\ket{\Psi_{\rm in}}$ is the joint state at the input of the interferometer, the expectation value of~$\hat{\mathcal{O}}= \ket{\Psi_{\rm in}}\bra{\Psi_{\rm in}}$ measured at the output 
$\ket{\Psi_{\rm out}} = \hat{U}^\dagger \rme^{-\rmi \hat{G}}\hat{U} \ket{\Psi_{\rm in}}$
of the interferometer is
\begin{equation}
\langle\hat{{\cal O}}\rangle_{\rm out} = \langle\Psi_{\rm out}|\hat{{\cal O}}|\Psi_{\rm out}\rangle = |\bra{\Psi_{\rm in}}\Psi_{\rm out}\rangle|^2 = 
|\bra{\Psi_{\rm in}} \hat{U}^\dagger \rme^{-\rmi \hat{G}}\hat{U} | \Psi_{\rm in}\rangle|^2  = |\bra{\Psi_U} \rme^{-\rmi \hat{G}} | \Psi_{U}\rangle|^2 = 
|\langle \rme^{-\rmi \hat{G}}\rangle_{U} |^2.
\end{equation}
By expanding the exponential we get
\begin{equation}
\label{ }
\langle \rme^{-\rmi \hat{G}}\rangle_{U} = \sum_{k\geq 0} \frac{(-\rmi)^k}{k!}  g^{(k)}, \qquad g^{(k)} = \langle \hat{G}^k \rangle_{U},
\end{equation}
whence
\begin{eqnarray}
\langle \hat{{\cal O}}\rangle_{\rm out}  &=& |\langle \rme^{-\rmi \hat{G}}\rangle_{U}|^2 = \sum_{j,k\geq 0} \frac{\rmi^j (-\rmi)^k}{j! k!}  g^{(j)} g^{(k)}
=  \sum_{j,k\geq 0} \frac{\rmi^j (-\rmi)^k}{j! k!}  g^{(j)} g^{(k)} \sum_{\ell\geq 0} \delta_{\ell, j+k}
= \sum_{\ell\geq 0} \sum_{0\leq k\leq \ell} \frac{\rmi^{\ell - k} (-\rmi)^k}{(\ell - k)! k!}  g^{(\ell -k)} g^{(k)} 
\nonumber\\
&=&  \sum_{\ell\geq 0} \frac{\rmi^\ell}{\ell!} \sum_{0\leq k\leq \ell} (-1)^k \frac{  \ell!}{k! (\ell - k)! } \,  g^{(\ell -k)} g^{(k)} , 
\end{eqnarray}
that is
\begin{equation}
\langle \hat{{\cal O}}\rangle_{\rm out} = \sum_{\ell\geq 0} \frac{\rmi^\ell}{\ell!} \mathcal{O}_\ell, \qquad 
\mathcal{O}_\ell = \sum_{0\leq k\leq \ell} (-1)^k \binom{\ell}{k} \,  g^{(\ell -k)} g^{(k)}.
\end{equation}

It is easy to show that $\mathcal{O}_\ell = 0$ for odd $\ell$:
\begin{eqnarray}
\mathcal{O}_{2m +1} &=& \sum_{0\leq k \leq m} (-1)^k \binom{2m+1}{k} \,  g^{(2m+1 -k)} g^{(k)}
+ \sum_{m+1\leq k \leq 2m+1} (-1)^k \binom{2m+1}{k} \,  g^{(2m+1 -k)} g^{(k)}
\nonumber\\
&=& \sum_{0\leq k \leq m} (-1)^k \binom{2m+1}{k} \,  g^{(2m+1 -k)} g^{(k)}
+ \sum_{0\leq k \leq m} (-1)^{2m+1- k} \binom{2m+1}{2m+1-k} \,  g^{(k)} g^{(2m+1-k)}
\nonumber\\
&=& \sum_{0\leq k \leq m} (-1)^k \binom{2m+1}{k} \,  g^{(2m+1 -k)} g^{(k)}
- \sum_{0\leq k \leq m} (-1)^{k} \binom{2m+1}{k} \,  g^{(k)} g^{(2m+1-k)} = 0,
\end{eqnarray}
and thus
\begin{equation}
\langle \hat{{\cal O}}\rangle_{\rm out} = \sum_{\ell\geq 0} \frac{(-1)^\ell}{(2\ell)!} \mathcal{O}_{2 \ell}.
\label{eq:seriesO}
\end{equation}

Notice that formally we have $\mathcal{O}_\ell = (g - g)^{(\ell)}$. In particular, we get
\begin{align}
\mathcal{O}_{0} &= g^{(0)} =1, 
\\
\mathcal{O}_{2} &= g^{(2)}  - 2 g^{(1)} g^{(1)} + g^{(2)}  = 2 \big(\langle \hat{G}^2 \rangle_{U} - \langle \hat{G} \rangle_{U}^2 \big),
\\
\mathcal{O}_{4} &= g^{(4)}  - 4 g^{(3)} g^{(1)} + 6 g^{(2)} g^{(2)} - 4 g^{(1)} g^{(3)} + g^{(4)} = 2 \big(\langle \hat{G}^4 \rangle_{U} 
- 4 \langle \hat{G}^3 \rangle_{U} \langle \hat{G} \rangle_{U}
+ 3 \langle \hat{G}^2 \rangle_{U}^2 \big),
\\
\mathcal{O}_{6} &= g^{(6)}  - 6 g^{(5)} g^{(1)} + 15 g^{(4)} g^{(2)} - 20 g^{(3)} g^{(3)} + \cdots = 2 \big(\langle \hat{G}^6 \rangle_{U} 
- 6 \langle \hat{G}^5 \rangle_{U} \langle \hat{G} \rangle_{U} + 15 \langle \hat{G}^4 \rangle_{U} \langle \hat{G}^2 \rangle_{U}
-10 \langle \hat{G}^3 \rangle_{U}^2 \big),
\end{align}
The 0-order term is just what we would obtain if there was no phase shift at all in any channel. The subsequent terms are corrections for small non-zero phases. 

Now we discuss the validity of the series expansion~\eqref{eq:seriesO}. 
Let 
\begin{equation}
|\varphi|_{\mathrm{max}} = \max_{1\leq j \leq M} |\varphi_j|.
\end{equation}. 
Given a state $\ket{\psi}$ we get that
\begin{equation}
\bra{\psi} \hat{G} \ket{\psi}= \sum_{j=1}^M \alpha_j \bra{\psi} \hat{a}_j^\dagger \hat{a}_j \ket{\psi} \leq \sum_{j=1}^M |\alpha_j| \bra{\psi} \hat{a}_j^\dagger \hat{a}_j \ket{\psi} \leq
|\varphi|_{\mathrm{max}}  \sum_{j=1}^M  \bra{\psi} \hat{a}_j^\dagger \hat{a}_j \ket{\psi} = |\varphi|_{\mathrm{max}} \bra{\psi}   \hat{N} \ket{\psi} ,
\end{equation}
since $\bra{\psi} \hat{a}_j^\dagger \hat{a}_j \ket{\psi}\geq 0$ for all $j$, whence we have the operator inequality
\begin{equation}
\hat{G} \leq  |\varphi|_{\mathrm{max}}  \hat{N}.
\end{equation}
By the same token one gets
\begin{equation}
\hat{G}  \geq  - |\varphi|_{\mathrm{max}}   \hat{N} ,
\end{equation}
and thus
\begin{equation}
|\bra{\psi} \hat{G}^k \ket{\psi}| \leq |\varphi|_{\mathrm{max}}^k \bra{\psi} \hat{N}^k \ket{\psi}
\end{equation}
for all states $\ket{\psi}$.
Since the input state $\ket{\Psi_{\mathrm{in}}}$ is Gaussian all moments of $\hat{N}$ are finite and one gets
\begin{equation}
\langle \hat{N}^k\rangle = \sum_{j=1}^kc_j \langle \hat{N} \rangle^j,
\end{equation}
for suitable $c_j \geq 0$. Therefore, for $\langle\hat{N}\rangle\ge1$, we have
\begin{equation}
\langle \hat{N}^k\rangle \leq c'_k \langle \hat{N} \rangle^k.
\end{equation}
Moreover, for any passive linear transformation $\hat{U}$, the total number $\hat{N}$ is conserved, and thus
\begin{equation}
\langle \hat{N}^k\rangle_U  = \langle \hat{N}^k\rangle.
\end{equation}
It follows that
\begin{equation}
|\langle \hat{G}^k\rangle_U| \leq c'_k \big( |\varphi|_{\mathrm{max}}   \langle   \hat{N}  \rangle \big)^k,
\end{equation}
so that for 
\begin{equation}
|\varphi|_{\mathrm{max}}   \langle   \hat{N} \rangle \ll 1,
\end{equation}
one can keep only the first nontrivial terms in the series expansion.

\subsection{Variance of the Generator of Phase Shifts}
\label{sec:variance_computation}

The expectation of $\hat{G}$ reads
\begin{equation}
\langle\hat{G} \rangle_U = \sum_j \varphi_j \langle \hat{a}^\dagger_j \hat{a}_j \rangle_U = \sum_j  \varphi_j  \bra{\Psi_{\mathrm{in}}} U^\dagger \hat{a}^\dagger_j \hat{a}_j U \ket{\Psi_{\mathrm{in}}}
= \sum_{j,l,m} \varphi_j\,  \mathcal{U}^*_{j l} \mathcal{U}_{jm} \langle \hat{a}^\dagger_l \hat{a}_m \rangle,
\end{equation}
where all sums range from 1 to $M$, and $\langle\hat{a}^\dagger_l \hat{a}_m \rangle = \bra{\Psi_{\mathrm{in}}} \hat{a}^\dagger_l \hat{a}_m \ket{\Psi_{\mathrm{in}}}$.
Since the only populated mode is the first, we have
\begin{equation}
\langle \hat{a}^\dagger_l \hat{a}_m \rangle = \delta_{l,1} \delta_{m,1} \langle \hat{a}^\dagger_1 \hat{a}_1 \rangle  = \delta_{l,1} \delta_{m,1} \langle\hat{N} \rangle,
\end{equation}
whence
\begin{equation}
\langle \hat{G} \rangle_U =  \sum_j \varphi_j |\mathcal{U}_{j1}|^2 \langle \hat{N} \rangle = \overline{\varphi}\, \langle \hat{N} \rangle,
\label{eq:Gav}
\end{equation}
where
\begin{equation}
\overline{f(\varphi)} =  \sum_j |\mathcal{U}_{j 1}|^2 f(\varphi_j).
\end{equation}

The expectation of $\hat{G}^2$ reads
\begin{equation}
\langle \hat{G}^2 \rangle_U = \sum_{j_1,j_2} \varphi_{j_1} \varphi_{j_2}  \langle \hat{a}^\dagger_{j_1} \hat{a}_{j_1}  \hat{a}^\dagger_{j_2} \hat{a}_{j_2} \rangle_U
= \sum_{j_1,j_2} \sum_{l_1,l_2} \sum_{m_1,m_2} \varphi_{j_1} \varphi_{j_2}  
\mathcal{U}^*_{j_1 l_1} \mathcal{U}_{j_1 m_1} \mathcal{U}^*_{j_2 l_2} \mathcal{U}_{j_2 m_2}
\langle \hat{a}^\dagger_{l_1} {\color{red}\hat{a}}_{m_1} \hat{a}^\dagger_{l_2} \hat{a}_{m_2} \rangle.
\end{equation}
By the canonical commutation relations,
\begin{equation}
\hat{a}^\dagger_{l_1} \hat{a}_{m_1} \hat{a}^\dagger_{l_2} \hat{a}_{m_2} = \hat{a}^\dagger_{l_1}  \hat{a}^\dagger_{l_2} \hat{a}_{m_1} \hat{a}_{m_2} + \delta_{m_1,l_2}\hat{a}^\dagger_{l_1} \hat{a}_{m_2} ,
\end{equation}
and thus
\begin{align}
\langle \hat{a}^\dagger_{l_1} \hat{a}_{m_1} \hat{a}^\dagger_{l_2} \hat{a}_{m_2} \rangle 
&= \langle \hat{a}^\dagger_{l_1} \hat{a}^\dagger_{l_2} \hat{a}_{m_1} \hat{a}_{m_2} \rangle + \delta_{m_1,l_2} \langle \hat{a}^\dagger_{l_1} \hat{a}_{m_2} \rangle
\notag\\
&= \delta_{l_1,1} \delta_{l_2,1} \delta_{m_1,1} \delta_{m_2,1} \langle \hat{a}^\dagger_{1}  \hat{a}^\dagger_{1} \hat{a}_{1} \hat{a}_{1} \rangle +
\delta_{m_1,l_2} \delta_{l_1,1} \delta_{m_2,1} \langle \hat{a}^\dagger_{1} \hat{a}_{1} \rangle
\notag\\
&= \delta_{l_1,1} \delta_{l_2,1} \delta_{m_1,1} \delta_{m_2,1} \big(\langle\hat{N}^2 \rangle -  \langle \hat{N} \rangle\big)  +
\delta_{m_1,l_2} \delta_{l_1,1} \delta_{m_2,1} \langle \hat{N} \rangle.
\end{align}
Therefore, we get
\begin{align}
\langle \hat{G}^2 \rangle_U &=  \sum_{j_1,j_2} \varphi_{j_1} \varphi_{j_2} |\mathcal{U}_{j_1 1}|^2 |\mathcal{U}_{j_2 1}|^2 \big(\langle \hat{N}^2 \rangle -  \langle \hat{N} \rangle\big) 
+ \sum_{j_1,j_2} \sum_{m_1} \varphi_{j_1} \varphi_{j_2}  \mathcal{U}^*_{j_1 1} \mathcal{U}_{j_1 m_1} \mathcal{U}^*_{j_2 m_1} \mathcal{U}_{j_2 1} \langle \hat{N} \rangle
\notag\\
&=  \sum_{j_1,j_2} \varphi_{j_1} \varphi_{j_2} |\mathcal{U}_{j_1 1}|^2 |\mathcal{U}_{j_2 1}|^2 \big(\langle \hat{N}^2 \rangle -  \langle \hat{N} \rangle\big) 
+ \sum_{j_1} \varphi_{j_1}^2 |\mathcal{U}_{j_1 1}|^2  \langle \hat{N} \rangle
\notag\\
&=  \overline{\varphi}^2 \, \big(\langle \hat{N}^2 \rangle -  \langle \hat{N} \rangle\big)  
+ \overline{\varphi^2} \, \langle \hat{N} \rangle,
\label{eq:G2av}
\end{align}
since $\sum_m \mathcal{U}_{j_1 m} \mathcal{U}^*_{j_2 m} = \delta_{j_1,j_2}$ by the unitarity of $\mathcal{U}$.

By gathering~\eqref{eq:G2av} and~\eqref{eq:Gav}, we finally get
\begin{align}
\Delta \hat{G}^2_U &= \langle \hat{G}^2 \rangle_U - \langle \hat{G} \rangle_U^2 =  \overline{\varphi}^2 \, \big(\langle \hat{N}^2 \rangle -  \langle \hat{N} \rangle\big) 
+ \overline{\varphi^2} \, \langle \hat{N} \rangle -  \overline{\varphi}^2\, \langle \hat{N} \rangle^2 
\notag\\
&= \overline{\varphi}^2 \, \bigl(\langle \hat{N}^2 \rangle -  \langle \hat{N} \rangle^2 \bigr) + \bigl(\overline{\varphi^2} - \overline{\varphi}^2\bigr)\, \langle \hat{N} \rangle ,
\end{align}
which is the sum of a contribution from  number fluctuations and a contribution from the fluctuations of the phases $\varphi_j$ according to the probability weights $|\mathcal{U}_{j 1}|^2$.

This result is valid for any $M$-boson state $\ket{\Psi_{\mathrm{in}}}$ with all modes but the first in the vacuum. Now let us use the fact that the first mode is in a squeezed vacuum state with mean photon number $\langle \hat{N} \rangle = \bar{N} = \sinh^2 |z|$ and mean squared photon number
\begin{align}
\langle \hat{N}^2\rangle &= \sum_{i,j=1}^M\bra{\Omega}\hat{S}^\dag_1(z)\hat{a}^\dag_i\hat{a}_i\hat{a}^\dag_j\hat{a}_i\hat{S}_1(z)\ket{\Omega} = \bra{0}\hat{S}^\dag_1(z)\hat{a}^\dag_1\hat{S}_1(z)\hat{S}^\dag_1(z)\hat{a}_1\hat{S}_1(z)\hat{S}^\dag_1(z)\hat{a}^\dag_1\hat{S}_1(z)\hat{S}^\dag_1(z)\hat{a}_1\hat{S}_1(z)\ket{0} = \nonumber\\
&= \sinh^2|z|\bra{0}\cosh^2|z|\hat{a}^2_1\hat{a}^{\dag2}_1 + \sinh^2|z|\hat{a}_1\hat{a}^\dag_1\hat{a}_1\hat{a}^\dag_1\ket{0} = \sinh^2|z|(2+3\sinh^2|z|),
\end{align}
where $\ket{\Omega}=\ket{0}_1\cdots\ket{0}_M$, and we have made use of the identity
\begin{equation}
 \hat{S}^\dag_i(z)\hat{a}_i\hat{S}_i(z) = \cosh|z|\,\hat{a}_i - \frac{z}{|z|}\sinh|z|\,\hat{a}_i^\dag,
\end{equation}
where $\hat{S}_i(z)=\rme^{\frac{1}{2}(z^*\hat{a}^2_i-z\hat{a}^{\dag2}_i)}$ is the squeezing operator.
The photon-number statistics in such a state is therefore super-Poissonian, with twice the variance of the Bose-Einstein distribution,
\begin{equation}
\Delta N^2 = \langle \hat{N}^2 \rangle -  \langle \hat{N} \rangle^2 = 2\sinh^2|z|(\sinh^2|z|+1) = 2\bar{N}\big(\bar{N}+1\big).
\label{eq:number_sq_var}
\end{equation} 
Hence, we get
\begin{equation}
\Delta G^2_U  = 2  \bar{N}\big(\bar{N}+1\big)\,  \overline{\varphi}^2  + \bar{N}\bigl(\overline{\varphi^2} - \overline{\varphi}^2\bigr).
\end{equation}
By taking only the dominant term in $\bar{N}$, we have
\begin{equation}
\Delta G^2_U  \simeq 2  \bar{N}^2 \,  \overline{\varphi}^2 = 2  \bar{N}^2 \, \Bigl(\sum_{j=1}^M |{\cal U}_{j 1}|^2 \varphi_j \Bigr)^2.
\end{equation}
\end{widetext}

\begin{thebibliography}{99}
\bibitem{nota1}
This sign restriction could possibly be lifted by occupying more than a single input channel, but this is out of the scope of this letter.
\end{thebibliography}
\end{document}